\documentclass[runningheads]{llncs}
\usepackage{amsmath,amssymb}
\usepackage{mathtools} 
\usepackage{tikz}
\usepackage{ltl}
\usepackage[ruled,vlined]{algorithm2e}
\usepackage[bordercolor=TodoRed, backgroundcolor=TodoRed!25, linecolor=TodoRed]{todonotes} 
\usepackage{color}
\usepackage{colortbl}
\usetikzlibrary{decorations.pathreplacing}
\usetikzlibrary{arrows,petri,backgrounds, positioning, shapes, patterns,calc,automata,intersections,calc}
\usepackage{wrapfig}
\usepackage{caption}
\usepackage{subcaption}
\usepackage{hyperref}
\usepackage{bm}

\newcommand{\cceq}{\mathop{::=}}

\renewcommand{\epsilon}{\varepsilon}
\renewcommand{\phi}{\varphi}

\newcommand{\pow}[1]{2^{#1}}
\newcommand{\nats}{\mathbb{N}}

\newcommand{\set}[1]{\{ #1 \}}
\newcommand{\ap}[0]{\mathrm{AP}}

\newcommand\equ{\leftrightarrow}
\newcommand{\ldot}{\mathpunct{.}}
\newcommand{\nat}{\mathbb{N}}

\newcommand{\hyltl}{{HyperLTL}\xspace}

\newcommand{\foe}{\mbox{FO[$<,E$]}\@\xspace}
\newcommand{\seins}{S1S[$E$]\@\xspace}
\newcommand{\msoe}{MSO[$E$]\@\xspace}
\newcommand{\mple}{MPL[$E$]\@\xspace}

\newcommand{\true}{\mathbf{tt}}
\newcommand{\false}{\mathbf{ff}}
\newcommand{\F}{{\mathbf{F\,}}}
\newcommand{\G}{{\mathbf{G\,}}}
\newcommand{\U}{{\mathbf{\,U\,}}}
\newcommand{\X}{{\mathbf{X\,}}}

\newcommand{\Waitfor}{\,\mathcal W\,}

\newcommand{\var}{\mathcal{V}}

\newcommand{\pathassign}{\Pi}

\newcommand{\Veins}{\mathcal{V}_1}
\newcommand{\Vzwei}{\mathcal{V}_2}

\newcommand{\Tra}{\ensuremath{\mathit{Tr}}}
\newcommand{\Paths}{\ensuremath{\mathit{Paths}}}

\newcommand{\donotshow}[1]{}

\begin{document}

\title{Model Checking Algorithms for Hyperproperties (invited paper)} 

\author{Bernd Finkbeiner \orcidID{0000-0002-4280-8441}}

\institute{CISPA Helmholtz Center for Information Security, Saarbr\"ucken, Germany\\
\email{finkbeiner@cispa.saarland}}

\maketitle
\begin{abstract}
 Hyperproperties generalize trace properties by expressing relations
 between multiple computations. Hyperpropertes include policies from
 information-flow security, like observational determinism or
 noninterference, and many other system properties including
 promptness and knowledge.  In this paper, we give an overview on the
 model checking problem for temporal hyperlogics.  Our starting point
 is the model checking algorithm for HyperLTL, a reduction to B\"uchi
 automata emptiness. This basic construction can be extended with
 propositional quantification, resulting in an algorithm for
 HyperQPTL. It can also be extended with branching time, resulting in
 an algorithm for HyperCTL$^*$.  However, it is not possible to have
 both extensions at the same time: the model checking problem of
 HyperQCTL$^*$ is undecidable.  An attractive compromise is offered
 by MPL[$E$], i.e., monadic path logic extended with the equal-level
 predicate. The expressiveness of MPL[$E$] falls strictly between that of
 HyperCTL$^*$ and HyperQCTL$^*$.   MPL[$E$] subsumes both HyperCTL$^*$ and
 HyperKCTL$^*$, the extension of HyperCTL$^*$ with the knowledge
 operator. We show that the model checking problem for MPL[$E$] is
 still decidable.

\end{abstract}

\section{Introduction}

 In recent years, the linear-time and branching-time temporal logics
 have been extended to allow for the specification of
 hyperproperties~\cite{DBLP:conf/post/ClarksonFKMRS14,Martin,journals/eatcs/Finkbeiner17,DBLP:conf/lics/CoenenFHH19,10.1007/978-3-030-59152-6_27}. Hyperproperties are a generalization of trace properties.
 Instead of properties of individual computations, hyperproperties
 express \emph{relations} between multiple computations~\cite{Hyperproperties}. This makes it
 possible to reason uniformly about system properties like information flow,
 promptness, and knowledge.

 In model checking, hyperproperties have played a significant role even
 before these new logics became available. An early insight was that
 the verification of a given system against properties
 that refer to multiple traces can be reduced to the verification of a
 \emph{modified} system against properties over individual
 traces. The idea is to self-compose the given system a sufficient number of times. The resulting traces  contain in each position a tuple of
 observations, each resulting from a different computation of the system.
 With this principle, certain
 hyperproperties like observational determinism and
 noninterference can be verified using model checking algorithms for standard
 linear and branching-time
 logics~\cite{HuismanWS/06/TLCharacterisationOfOD,DBLP:journals/mscs/BartheDR11,VanDerMeyden:2007:verifOfNonInterf}.

 The development of new logics specifically for hyperproperties
 considerably broadened the range of hyperproperties that can be checked automatically. HyperLTL is an extension of linear-time temporal logic (LTL) with quantifiers over trace variables, which allow the formula to refer to multiple traces at the same time. For example, \emph{noninterference}~\cite{Goguen+Meseguer/1982/SecurityPoliciesAndSecurityModels} between a secret input $h$ and a public output $o$ can be specified in HyperLTL by requiring that all pairs of traces $\pi$ and $\pi'$ that have, in every step, the same inputs except for $h$ (i.e., all inputs in $I\setminus\{h\}$ are equal on $\pi$ and $\pi'$) also have the same output $o$ at all times:
\[
\forall\pi.\forall\pi'.~ \G \big(\!\!\!\bigwedge_{i\in I\setminus \{h\}}\! i_\pi = i_{\pi'}\big) ~\Rightarrow~ \G\, (o_\pi = o_{\pi'})
\]
By combining universal and existential quantification, HyperLTL can also express properties like \emph{generalized noninterference} (GNI)~\cite{McCullough:1987:GNI}, which requires that for  every pair of traces $\pi$ and $\pi'$, there is a third trace $\pi''$ that agrees with $\pi$ on $h$ and with $\pi'$ on $o$:
\[
\forall\pi.\forall\pi'.\exists\pi''.~\G\, (h_\pi = h_{\pi''}) ~\wedge~ \G\, (o_{\pi'} = o_{\pi''})
\]

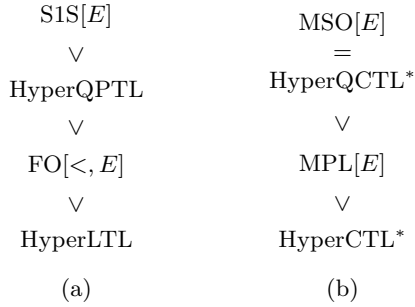
\begin{figure}[t]
  \centering
\begin{subfigure}[t]{.29\columnwidth}
	\centering
	\begin{tikzpicture}[every node/.append style = {font=\small}]
		\node (f) at (0,3) {\seins};
		\node (g) at (0,2) {HyperQPTL};
		\node (h) at (0,1) {\foe};
		\node (j) at (0,0) {HyperLTL};
		
		\draw[-, color = white] (j) edge node[sloped, color = black] {$<$} (h)
				(h) edge node[sloped, color = black] {$<$} (g)
				 (g) edge node[sloped, color = black] {$<$} (f);
	\node (z) at (0.7,2.5) {};
	\node (z) at (0.7,1.5) {};
	\node (z) at (0.7,0.5) {};
	\end{tikzpicture}
	\caption{}
	\label{fig:lin_hyper}
\end{subfigure}
\begin{subfigure}[t]{.29\columnwidth}
	\centering
	\begin{tikzpicture}[every node/.append style = {font=\small}]
          \node[color=white] (h) at (0,3) {MSO[E]};
		\node[align=center] (h) at (0,2.5) {\msoe \\ = \\ HyperQCTL$^*$};
		\node (i) at (0,1) {\mple};
		\node (j) at (0,0) {HyperCTL$^*$};
		
		\draw[-, color = white] (j) edge node[sloped, color = black] {$<$} (i)
				 (i) edge node[sloped, color = black] {$<$} (h);
	\node (z) at (0.7,2.5) {};
	\node (z) at (0.7,1.5) {};
	\node (z) at (0.7,0.5) {};
	\end{tikzpicture}
	\caption{}
	\label{fig:bran_hyper}
\end{subfigure}
\caption{The hierarchy of hyperlogics~\cite{DBLP:conf/lics/CoenenFHH19}: (a) linear time, (b) branching time.}
\label{fig:hier}
\end{figure}

HyperLTL is the starting point of an entire hierarchy of \emph{hyperlogics},
depicted in Fig.~\ref{fig:hier} and analyzed in detail in~\cite{DBLP:conf/lics/CoenenFHH19}. The hyperlogics are obtained from their classic counterparts with two principal extensions. The temporal logics LTL, QPTL, and CTL$^*$ are extended with quantifiers and variables over traces or paths, such that the formula can refer to multiple traces or paths at the same time; the first-order and second-order logics FO, S1S, MPL, and MSO are extended with the equal-level predicate $E$, which indicates that two points happen at the same time (albeit possibly on different computations of the system).

A key limitation of HyperLTL, as first pointed out by Bozzelli et al.~\cite{BozzelliMP15}, is that it is not possible to express promptness requirements, which say that there should exist a common deadline over all traces by which a certain eventuality is satisfied. Such properties can be expressed in \foe, monadic first-order logic of order extended with the equal-level predicate. \foe is subsumed by the temporal logic HyperQPTL, which extends HyperLTL with quantification over propositions. 
The following HyperQPTL formula specifies the existence of a common deadline over all traces by which a certain predicate $p$ must become true on all traces. The quantification over the proposition $d$, which expresses the common deadline, introduces a valuation of $d$ that is \emph{independent} of the choice of trace $\pi$:
\[
\boldsymbol{\exists} d \ldot \forall \pi \ldot  \neg d\, \LTLuntil\, (p_{\pi} \land \F d)
\]
HyperQPTL captures the \emph{$\omega$-regular hyperproperties}~\cite{10.1007/978-3-030-53291-8_4}. Even more expressive is \seins, monadic second order logic with one successor equipped with the equal-level predicate. While the model checking problem of HyperQPTL is still decidable, it becomes undecidable for \seins. This is different from the case of trace properties, where S1S is equally expressive to QPTL, and both have decidable model checking problems.

Extending HyperLTL to branching time leads to the temporal logic Hyper\-CTL$^*$~\cite{DBLP:conf/post/ClarksonFKMRS14}, which has the same syntax as HyperLTL, except that the quantifiers refer to paths, rather than traces, and that path quantifiers may occur in the scope of temporal modalities. HyperCTL$^*$ is subsumed by monadic path logic equipped with the equal-level predicate (\mple), which is a second-order logic where second-order quantifiers are restricted to full computation paths.
\mple in turn is contained in HyperQCTL$^*$, the extension of HyperCTL$^*$ with propositional quantification. HyperQCTL$^*$ is as expressive as full monadic second-order logic with the equal-level predicate (\msoe)~\cite{DBLP:conf/lics/CoenenFHH19}.

In this paper, we study this hierarchy of logics from the perspective of the model checking problem. Our starting point is the model checking algorithm for HyperLTL, which reduces the model checking problem to the language emptiness problem of a B\"uchi automaton~\cite{DBLP:conf/cav/FinkbeinerRS15}. The construction is similar to the idea of self-composition in that for every trace variable a separate copy of the system is introduced. Quantifiers are then eliminated by existential and universal projection on the language of the automaton. 
This basic construction can be extended with propositional quantification, which is also handled by projection. The construction can also be extended to branching time, by tracking the precise state of each computation, rather than just the trace label. However, it is not possible to implement both extensions at the same time: the model checking problem of HyperQCTL$^*$ is undecidable~\cite{DBLP:conf/lics/CoenenFHH19}.

\donotshow{ 
The intuitive reason for the undecidability is that the propositional quantification allows us to label the computation tree and, hence, identify arbitrary subsets of the computations. In this way, we can encode the satisfiability problem of HyperLTL, which is undecidable, as a model checking problem for a HyperQCTL$^*$ formula.
}

The undecidability of HyperQCTL$^*$ is unfortunate, because many interesting properties, such as branching-time knowledge, can be expressed in HyperQCTL$^*$, but not in HyperCTL$^*$. It turns out, however, that \mple, whose expressiveness lies strictly between HyperCTL$^*$ and HyperQCTL$^*$, still has a decidable model checking problem. As the only original contribution of this paper (everything else is based on previously published results), we present the first model checking algorithm for \mple. MPL[$E$] is a very attractive compromise. MPL[$E$] subsumes both HyperCTL$^*$ and HyperKCTL$^*$~\cite{DBLP:conf/lics/CoenenFHH19}, the extension of HyperCTL$^*$ with the knowledge operator.

\section{HyperLTL}

HyperLTL is a generalization of linear-time temporal logic (LTL). We quickly review the syntax and semantics of LTL and then describe the extension to HyperLTL.
Let $\ap$ be a finite set of atomic propositions. A trace over $\ap$ is a map $t \colon \nats \rightarrow \pow{\ap}$, denoted by $t(0)t(1)t(2) \cdots$. Let $(\pow{\ap})^\omega$ denote the set of all traces over $\ap$.

\paragraph{LTL.}
The formulas of linear-time temporal logic (LTL)~\cite{Pnueli/1977/TheTemporalLogicOfPrograms} 
are generated by the following grammar: \vspace{-1mm}
$$
\varphi~~::=~~ a ~~|~~ \neg\varphi ~~|~~ \varphi\wedge\varphi ~~|~~ \X \varphi ~~|~~ \varphi \U \varphi \vspace{-.5mm}
$$
where $a\in\ap$ is an \emph{atomic proposition}, the Boolean connectives $\neg$ and $\wedge$ have the usual meaning, $\X$ is the temporal \emph{next} operator, and $\U$ is the temporal \emph{until} operator. 
We also consider the usual derived Boolean connectives, such as $\vee$, $\rightarrow$, and $\equ$, and the derived temporal operators \emph{eventually} $\F\varphi\equiv \true\U\varphi$, \emph{globally} $\G\varphi\equiv\neg\F\neg\varphi$, and \emph{weak until}: $\varphi \Waitfor \psi \equiv (\varphi \U \psi) \vee \G \varphi$.
The satisfaction of an LTL formula $\varphi$ over a trace $t$ at a position $i\in \mathbb N$, denoted by $t,i \models \varphi$, is defined as follows:
\[
\begin{array}{l@{\hspace{1em}}c@{\hspace{1em}}l}
  t,i\models a & \text{iff} &  a\in t(i),\\
  t,i \models \neg \varphi & \text{iff}&  t,i \not\models \varphi,\\
  t,i\models \varphi_1 \wedge \varphi_2 & \text{iff} &  t,i \models \varphi_1 \text{ and } t,i \models \varphi_2,\\
  t,i\models\X\varphi & \text{iff} &  t,i+1\models\varphi,\\
  t,i\models \varphi_1\!\U\!\varphi_2 & \text{iff} & \exists k \geq i :~t,k \models\varphi_2~\wedge~\forall i \leq j < k:~t,j \models\varphi_1.
\end{array}
\]
We say that a trace $t$ satisfies a sentence~$\phi$, denoted by $t \models \phi$, if $t,0 \models \phi$.
For example, the LTL formula $\G(a\rightarrow\F b)$ specifies that every position in which $a$ is true must eventually be followed by a position where $b$ is true.


\paragraph{HyperLTL.}
The formulas of \hyltl~\cite{DBLP:conf/post/ClarksonFKMRS14} are generated by the grammar 
\begin{align*}
\phi & {} \cceq {}  \exists \pi.\ \phi ~\mid~ \forall \pi.\ \phi ~\mid~ \psi \\
\psi & {}  \cceq {}  a_\pi ~\mid~ \neg \psi ~\mid~ \psi \wedge \psi ~\mid~ \X \psi ~\mid~ \psi \U \psi 
\end{align*}
where $a$ is an atomic proposition from a set $\ap$ and $\pi$ is a trace variable from a set $\var$. Further Boolean connectives and the temporal operators $\F$, $\G$, and $\mathcal W$ are derived as for LTL. A sentence is a closed formula, i.e., the formula has no free trace variables.

The semantics of \hyltl is defined with respect to a trace assignment, a partial mapping~$\Pi \colon \var \rightarrow (\pow{\ap})^\omega$. The assignment with empty domain is denoted by $\Pi_\emptyset$. Given a trace assignment~$\Pi$, a trace variable~$\pi$, and a trace~$t$, we denote by $\Pi[\pi \rightarrow t]$ the assignment that coincides with $\Pi$ everywhere but at $\pi$, which is mapped to $t$.
%
The satisfaction of a HyperLTL formula $\varphi$ over a trace assignment $\Pi$ and a set of traces $T$ at a position $i \in \mathbb N$, denoted by $T,\Pi,i \models \varphi$, is defined as follows:

\[
\begin{array}{l@{\hspace{1em}}c@{\hspace{1em}}l}
  T, \Pi,i \models a_\pi & \text{iff} & a \in \Pi(\pi)(i),\\
  T, \Pi,i \models \neg \psi & \text{iff} & T, \Pi,i \not\models \psi,\\
  T, \Pi,i \models \psi_1 \wedge \psi_2 & \text{iff} & T, \Pi,i \models \psi_1 \text{ and } T, \Pi,i \models \psi_2,\\
  T, \Pi,i \models \X \psi & \mbox{iff} & T,\Pi,i+1 \models \psi,\\
  T, \Pi,i \models \psi_1 \U \psi_2 & \text{iff} &  \exists k \ge i: T,\Pi, k \models \psi_2\\
  & & \wedge \forall i \le j < k: T,\Pi,j \models \psi_1,\\
  T, \Pi,i \models \exists \pi.\ \phi & \text{iff} & \exists t \in T: T,\Pi[\pi \rightarrow t],i \models \psi,\\
  T, \Pi,i \models \forall \pi.\ \phi & \text{iff} & \forall t \in T: T,\Pi[\pi \rightarrow t],i \models \psi.
  \end{array}
\]

We say that a set $T$ of traces satisfies a sentence~$\phi$, denoted by $T \models \phi$, if $T, \Pi_\emptyset,0 \models \phi$.

\paragraph{System properties.}
	A \emph{Kripke structure} is a tuple $K=(S,s_0,\delta,\ap,L)$ consisting of a set of states $S$, an initial state $s_0$, a transition function $\delta:S\to 2^{S}$, a set of \emph{atomic propositions} $\ap$, and a \emph{labeling function}  $L:S^*\to 2^{\ap}$ that assigns a set of atomic propositions that are true after a given sequence of states has been traversed.
	We require that each state has a successor, that is $\delta(s)\neq\emptyset$, to ensure that every execution of a Kripke structure can always be continued to infinity.
        In a \emph{finite} Kripke structure, $S$ is a finite set. We furthermore assume that in a finite Kripke structure, $L$ only depends on the last state, so that $L$ can also be given as a function $S \to 2^{\ap}$.

A \emph{path} of a Kripke structure is an infinite sequence $s_0s_1\ldots\in S^\omega$ such that $s_0$ is the initial state of $K$ and $s_{i+1}\in\delta(s_i)$ for all $i\in \mathbb{N}$. 
By $\Paths(K,s)$, we denote the set of all paths of $K$ starting in state $s\in S$.
  A \emph{trace} of a path $\sigma=s_0s_1\ldots$ is a sequence of labels $l_0l_1\ldots$ with $l_i=L(s_0s_1\ldots s_i)$ for all $i\in\mathbb{N}$. $\Tra(K,s)$ is the set of all traces of paths of a Kripke structure $K$ starting in state~$s$. 
  A Kripke structure $K$ with initial state $s_0$ satisfies an LTL formula $\varphi$, denoted by $K \models \varphi$ iff for all traces $\pi \in \Tra(K,s_0)$, it holds that $\pi \models \varphi$. Likewise, the Kripke structure satisfies a HyperLTL formula $\varphi$, also denoted by $K \models \varphi$, iff $\Tra(K,s_0) \models \varphi$.

\paragraph{Model checking.}
\label{section:modelchecking}
The HyperLTL model checking problem is to decide, for a given finite Kripke structure $K$ and a given HyperLTL formula $\psi$, whether or not $K \models \psi$. The following basic construction (described in more detail in~\cite{DBLP:conf/cav/FinkbeinerRS15}) reduces the model checking problem to the language emptiness problem of a B\"uchi automaton: the given Kripke structure satisfies the formula if and only if the language of the resulting automaton is empty. 

The construction starts by negating $\psi$, so that it describes the existence of an error. Since we assume that a HyperLTL formula begins with a quantifier prefix, this means that we dualize the quantifiers and then negate the inner LTL formula. Let us assume that the resulting HyperLTL formula has the form $Q_n \pi_{n-1} Q_2 \pi_{n-1}\ldots Q_1 \pi_1 .\ \varphi$ where $Q_1, Q_2, \ldots Q_n$ are trace quantifiers in $\{ \exists, \forall \}$ and $\varphi$ is a quantifier-free formula over atomic propositions indexed by trace variables $\{\pi_1, \ldots \pi_n\}$. 

Similar to standard LTL model checking, we convert the LTL formula $\varphi$  into an equivalent B\"uchi automaton $\mathcal A_0$ over the alphabet $(2^\ap)^n$. Each letter is a tuple of $n$ sets of atomic propositions, where the $i$th element of the tuple represents the atomic propositions of trace $\pi_i$.

  Next, the algorithm eliminates the quantifiers. For this purpose, it carries out $n$ steps that each eliminate one component from the tuple of the input alphabet. In the $i$th step, we eliminate the $i$th component, corresponding to trace variable~$\pi_i$.
  Let us consider the $i$th step.
  Over the previous steps, the automaton $\mathcal A_{i-1}$ over alphabet $(2^\ap)^{(n-i)}$ has been constructed, and now the first component of the tuple corresponds to $\pi_i$.  If the trace quantifier $Q_i$ is existential,
  we intersect $\mathcal A_{i-1}$ with the Kripke structure $K$ so that, in the sequence of letters, the first component of the tuple is chosen consistently with some path in $K$. Subsequently, we eliminate the first component of the tuple by existential projection on the automaton.
  If $Q_i$ is universal, then we combine $\mathcal A_{i-1}$ with the Kripke structure $K$ so that only sequences in which the first component is chosen consistently with some path in $K$ need to be accepted by $\mathcal A_{i-1}$. Subsequently, we eliminate the first component of the tuple by universal projection on the automaton. This results in the next automaton $\mathcal A_i$. 
  
After $n$ such steps, all quantifiers have been eliminated and the language of the resulting automaton is over the one-letter alphabet (consisting of the empty tuple). The HyperLTL formula is satisfied if and only if the language of automaton $\mathcal A_n$ is empty.

\donotshow{
  This construction can be optimized a little bit by considering multiple quantifiers  together if they appear next to each other in the quantifier prefix and have the same polarity. If, for example, two universal quantifiers are in subsequent positions of the quantifier prefix, then it is not necessary to complement the intermediate automaton after the first quantifier has been eliminated. If the polarity of the quantifiers changes, however, a complementation is unavoidable. This is reflected in the complexity of the model checking problem, where each quantifier alternation increases the complexity exponentially.
  
\begin{theorem}\cite{DBLP:conf/cav/FinkbeinerRS15,markusPhD}
Given a Kripke structure $K$ and a HyperCTL$^*$ formula $\varphi$ with
alternation depth $k$, the model checking problem for $K$ and $\varphi$ is complete for NSPACE$(g(k,|\varphi|))$ and NSPACE$(g(k-1,|K|))$.
\end{theorem}

As a result, hyperproperties that require no quantifier alternation have fairly inexpensive model checking problems. Observational determinism, for example, which is a universal property, is in NLOGSPACE in the size of the Kripke structure. Nininference and generalized noninterference, which bith have one quantifier alternation, are in PSPACE in the size of the Kripke structure.
}

\section{HyperQPTL}

HyperQPTL~\cite{markusPhD,DBLP:conf/lics/CoenenFHH19} extends HyperLTL with quantification over atomic propositions. To easily distinguish quantification over traces $\exists \pi, \forall \pi$ and quantification over propositions $\boldsymbol{\exists} p, {\boldsymbol{\forall}} p$, we use boldface for the latter. The formulas of HyperQPTL are generated by the following grammar:
\begin{align*}
\phi & {} \cceq {}  \exists \pi.\ \phi ~\mid~ \forall \pi.\ \phi ~\mid~ \psi ~\mid~ \boldsymbol{\exists} p.\ \phi ~\mid~ \boldsymbol{\forall} p.\ \phi ~\mid~ \psi \\
\psi & {}  \cceq {}  a_\pi ~\mid~ p ~\mid~ \neg \psi ~\mid~ \psi \wedge \psi ~\mid~ \X \psi ~\mid~ \F \psi 
\end{align*}
where $a,p \in \ap$ and $\pi \in \var$.
	The semantics of HyperQPTL corresponds to the semantics of HyperLTL with additional rules for propositional quantification:
	\begin{alignat*}{3}
	  &T, \pathassign,i  \models \boldsymbol{\exists} q \ldot \varphi \quad &&\text{ iff } && \quad \exists t \in (2^{\set{q}})^\omega.~ T, \pathassign[\pi_q \mapsto t],i  \models \varphi \\
          &T, \pathassign,i  \models \boldsymbol{\forall} q \ldot \varphi \quad &&\text{ iff } && \quad \forall t \in (2^{\set{q}})^\omega.~ T, \pathassign[\pi_q \mapsto t],i  \models \varphi \\
	&T, \pathassign, i \models q  \quad &&\text{ iff } && \quad q \in \pathassign(\pi_q)(i). 
	\end{alignat*}

\paragraph{Expressiveness.}\label{section:knowledgeop}
As discussed in the introduction, HyperQPTL can express \emph{promptness}~\cite{journals/fmsd/KupfermanPV09}, which states that there is a bound, common for all traces, until which an eventuality has to be fulfilled. Another common type of property that can be expressed in HyperQPTL is \emph{knowledge}. 
Epistemic temporal logics extend temporal logics with a so-called \emph{knowledge} operator $\mathcal{K}_A \varphi$, denoting that an agent $A$ knows $\varphi$.
HyperQPTL can be extended to HyperQPTL$_K$ as follows~\cite{markusPhD}:
\begin{alignat*}{3}
&T, \pathassign, i \models \mathcal{K}_{A,\pi} \varphi \quad &&\text{iff} && \quad \forall t' \in T \ldot \pathassign(\pi)[0, i] =_A t'[0, i] \rightarrow T, \pathassign[\pi \mapsto t'], i \models \varphi
\end{alignat*}
In this definition, $t[0,i]$ denotes the prefix of a trace $t$ up to position $i$. Two sequences $t,t'$ are equivalent with respect to agent $A$, denoted by $t =A t'$, if $A$ cannot distinguish $t$ and $t'$. We assume that $A$ is given as a set of atomic propositions $A \subseteq \ap$. Then $t =_A t'$ holds if $t$ and $t'$ agree on all propositions in~$A$.

As shown in \cite{markusPhD}, the knowledge operator can be eliminated, resulting in an equivalent HyperQPTL formula. The idea is to replace an application of the knowledge operator $\mathcal{K}_{A,\pi} \psi$  with an existentially quantified proposition $u$ and add the following requirement to ensure that $u$ is only true at positions where the knowledge formula is satisfied:
	\begin{align*}
		\forall r \ldot \forall \pi' \ldot 
		  ((r~ \LTLuntil ~(u\wedge r \wedge \LTLnext \LTLsquare \neg r)) \wedge \LTLsquare (r \rightarrow A_\pi = A_{\pi'}) \rightarrow \LTLsquare (r \wedge \LTLnext \neg r \rightarrow \psi[\pi' \slash \pi]))
	\end{align*}
In this definition, $A_\pi = A_{\pi'}$ is an abbreviation for the conjunction over all propositions in $A$ that ensures that each proposition has the same value in $\pi$ and in $\pi'$. 
For each position where the knowledge formula is claimed to be true, the universally quantified proposition $r$ changes from true to false at exactly that position, thus marking the prefix leading to this point. The knowledge formula is then true iff $\psi$ holds on all traces $\pi'$ that agree with respect to $A$ on the prefix.

\label{section:foe}
HyperQPTL is also strictly more expressive than \foe, the extension of the first-order logic of order with the equal-level predicate $E$~\cite{DBLP:conf/lics/CoenenFHH19}.
Given a set $V_1$ of first-order variables, the formulas $\varphi$ of \foe
are generated by the following grammar~\cite{Martin}:
\begin{align*}	
  \varphi &\Coloneqq \psi ~\mid~ \neg \varphi ~\mid~ \varphi_1 \vee \varphi_2  ~\mid~ \exists x. \varphi \\
  \psi  &\Coloneqq P_a(x) ~\mid~ x<y ~\mid~ x=y ~\mid~ E(x,y),
\end{align*}
where $a \in AP$ and $x, y \in V_1$.
We interpret \foe formulas over a set of traces $T$. 
We assign first-order variables to elements from the domain $T \times \mathbb N$.
We define the satisfaction relation $T, \Veins \models \varphi $ with respect to a valuation $\mathcal V_1$ assigning all free variables in $V'_1$ as follows:
\begin{alignat*}{3}
T, \Veins &\models P_a(x)              &&\text{ iff } &&a \in t(n) \mbox{ where } (t,n) = \Veins(x)  \\
T, \Veins &\models x<y               &&\text{ iff } && t_1 = t_2 \land n_1 < n_2 \mbox{ where } (t_1,n_1) = \Veins(x) \text{ and } (t_2,n_2) = \Veins(y) \\
T, \Veins &\models x=y               &&\text{ iff } &&  \Veins(x) = \Veins(y) \\
T, \Veins &\models E(x,y)               &&\text{ iff } && n_1 = n_2 \mbox{ where } (t_1,n_1) = \Veins(x) \text{ and } (t_2,n_2) = \Veins(y) \\
T, \Veins  &\models \neg \varphi         &&\text{ iff } &&T, \Veins \not\models \varphi \\
T, \Veins &\models \varphi_1 \vee \varphi_2 &&\text{ iff } &&T, \Veins \models \varphi_1 \text{ or } T, \Veins \models \varphi_2 \\
T, \Veins &\models \exists x. \varphi   &&\text{ iff } && \exists (t, n) \in T \times \nat.~ \\
& && && T, \Veins[x \mapsto (t, n)] \models \varphi,
\end{alignat*}
where $\mathcal{V}_1[x \mapsto v]$ updates a valuation. 
A trace set $T$ satisfies a closed \foe formula $\varphi$, written $T \models \varphi$, if $T, \emptyset \models \varphi$, where $\emptyset$ denotes the empty valuation.

\paragraph{Model checking.} The only required modification to the model checking algorithm described in Section~\ref{section:modelchecking} is the treatment of the propositional quantifiers. Since the valuation of the propositions is not restricted by the given Kripke structure, we omit the intersection with the Kripke structure for quantified propositions, and instead eliminate the quantifier by existential or universal projection only.

\section{Beyond HyperQPTL}

The model checking problems of linear-time hyperlogics beyond HyperQPTL quickly become undecidable. Two examples of such logics are HyperQPTL$^+$ and \seins.

\paragraph{HyperQPTL$^+$.} HyperQPTL$^+$~\cite{10.1007/978-3-030-53291-8_4} differs from HyperQPTL in the role of the propositional quantification. Rather than interpreting the quantified propositions with an additional sequence of values, HyperQPTL$^+$ modifies the interpretation on the existing traces. The syntax of HyperQPTL$^+$ is thus slightly simpler, because also the quantified propositions appear indexed with trace variables:
\begin{align*}
			\varphi &{}\Coloneqq \forall\pi\ldot\varphi ~\mid~ \exists\pi\ldot\varphi ~\mid~ \forall a \ldot\varphi ~\mid~ \exists a \ldot\varphi ~\mid~ \psi \enspace \\
			\psi &{}\Coloneqq a_\pi ~\mid~ \neg\psi ~\mid~ \psi\lor\psi ~\mid~ \X\psi ~\mid~ \F\psi \enspace .
		\end{align*}
		In the semantics, the rules for propositional quantification are changed accordingly:
		\begin{alignat*}{3}
		&T, \pathassign,i \models \exists a \ldot \varphi \quad && \text{ iff } \quad && \exists T' \subseteq (\pow{\ap})^\omega \ldot T' =_{\ap \backslash \{a\}} T \land T', \pathassign, i \models \varphi\\
		&T, \pathassign,i \models \forall a \ldot \varphi && \text{ iff } && \forall T' \subseteq (\pow{\ap})^\omega \ldot T' =_{\ap \backslash \{a\}} T \rightarrow T', \pathassign, i \models \varphi \enspace .
		\end{alignat*}

\paragraph{\seins.} \seins is  monadic second-order logic with one successor (S1S) extended with the equal-level predicate.
Let $V_1 = \{x_1, x_2, \ldots\}$ be a set of first-order variables, and $V_2 = \{X_1, X_2, \ldots\}$ a set of second-order variables.
The formulas $\varphi$ of \seins are generated by the following grammar:
\begin{align*}
\tau &\Coloneqq x ~\mid~ \mathit{min}(x) ~\mid~ \mathit{Succ}(\tau)\\
\varphi &\Coloneqq \tau \in X ~\mid~ \tau = \tau ~\mid~ E(\tau,\tau) ~\mid~ \neg \varphi ~\mid~ \varphi \vee \varphi ~\mid~ \exists x. \varphi ~\mid~ \exists X. \varphi,
\end{align*}
where $x \in V_1$ is a first-order variable, $\mathit{Succ}$ denotes the successor relation, and $\mathit{min}(x)$ indicates the minimal element of the traces addressed by $x$.
Furthermore, $E(\tau, \tau)$ is the equal-level predicate and $X \in V_2 \cup \{X_a ~|~ a \in AP \}$.
%
We interpret \seins formulas over a set of traces $T$. 
As for \foe, the domain of the first-order variables is $T \times \nat$.
Let $\mathcal{V}_1: V_1 \to T \times \mathbb{N}$ and $\mathcal{V}_2: V_2 \to 2^{(T\times \mathbb{N})}$ be the first-order and second-order valuation, respectively. 
The value of a term is defined as follows:
\begin{align*}
[x]_{\Veins} &= \Veins(x)\\
[\mathit{min}(x)]_{\Veins} &= (\mathit{proj}_1 (\mathcal{V}_1(x)),0)\\
[S(\tau)]_{\Veins} &= (\mathit{proj}_1([\tau]_{\Veins}) , \mathit{proj}_2([\tau]_{\Veins}) +1),
\end{align*}
where $\mathit{proj}_1$ and $\mathit{proj}_2$ denote the projection to the first and second component, respectively.
Let $\varphi$ be an \seins formula with free first-order and second-order variables $V'_1 \subseteq V_1$ and $V'_2 \subseteq V_2 \cup \{X_a ~|~ a \in AP \}$, respectively.
We define the satisfaction relation $T, \Veins, \Vzwei \models \varphi $ with respect to two valuations $\mathcal V_1, \mathcal V_2$ assigning all free variables in $V'_1$ and $V'_2$ as follows:
\begin{alignat*}{3}
T, \Veins, \Vzwei &\models \tau \in X              &&\text{ iff } &&[\tau]_{\Veins} \in \Vzwei(X) \\
T, \Veins, \Vzwei &\models \tau_1 =\tau_2                &&\text{ iff } &&[\tau_1]_{\Veins} = [\tau_2]_{\Veins} \\
T, \Veins, \Vzwei &\models E(\tau_1,\tau_2)               &&\text{ iff } &&\mathit{proj}_2([\tau_1]_{\Veins}) = \mathit{proj}_2([\tau_2]_{\Veins}) \\
T, \Veins, \Vzwei &\models \neg \varphi         &&\text{ iff } &&T, \Veins, \Vzwei \not\models \varphi \\
T, \Veins, \Vzwei &\models \varphi_1 \vee \varphi_2 &&\text{ iff } &&T, \Veins, \Vzwei \models \varphi_1 \text{ or } T, \Veins, \Vzwei \models \varphi_2 \\
T, \Veins, \Vzwei &\models \exists x. \varphi   &&\text{ iff } && \exists (t, n) \in T \times \nat.~ \\
& && && T, \Veins[x \mapsto (t, n)], \Vzwei \models \varphi \\
T, \Veins, \Vzwei &\models \exists X. \varphi   &&\text{ iff } && \exists A \subseteq T \times \nat.~ \\
& && && T, \Veins, \Vzwei[X \mapsto A] \models \varphi,
\end{alignat*}
where $\mathcal{V}_i[x \mapsto v]$ updates a valuation. 
A trace set $T$ satisfies a closed \seins formula $\varphi$, written $T \models \varphi$, if $T, \emptyset, \mathcal{V}_2 \models \varphi$, where $\emptyset$ denotes the empty first-order valuation and $\mathcal{V}_2$ assigns each free $X_a$ in $\varphi$ to the set $\{(t,n) \in T \times \nat~|~ a \in t[n]\}$.
                
\paragraph{Model checking.} The model checking problems of HyperQPTL$^+$ and \seins are both undecidable, as shown in \cite{10.1007/978-3-030-53291-8_4} and \cite{DBLP:conf/lics/CoenenFHH19}, respectively.

\section{HyperCTL$^*$}

Extending the path quantifiers of CTL$^*$ by \emph{path variables} leads to the logic HyperCTL$^*$, which subsumes both HyperLTL and CTL$^*$.
The formulas of Hyper\-CTL$^*$ are generated by the following grammar:
\[
\begin{array}{llllllllllllllllll}
\varphi ~&~::= ~&~ a_{\pi} ~&~ \vert ~&~ \neg\varphi ~&~ \vert ~&~ 
\varphi\vee\varphi ~&~ \vert ~&~ \LTLcircle\varphi ~&~ \vert ~&~ \varphi \U\varphi ~&~ \vert ~&~ \exists \pi.\; \varphi ~&~ 
\end{array}
\]
We require that temporal operators only occur inside the scope of path quantifiers. 
The semantics of HyperCTL$^*$ is given in terms of assignments of variables to \emph{paths}, which are defined analogously to trace assignments. 
Given a Kripke structure $K$, the satisfaction of a HyperCTL$^*$ formula $\varphi$ at a position $i \in \mathbb N$, denoted by $K, \Pi,i \models \varphi$, is defined as follows: 
\[
\begin{array}{l@{\hspace{1em}}c@{\hspace{1em}}l}
  K,\Pi,i\models a_{\pi} & \text{iff} & a\in L\big(\Pi(\pi)[0\ldots i]\big), \\
  K,\Pi,i\models \neg \varphi & \text{iff} & \Pi,K,i\not\models \varphi, \\
  K,\Pi,i\models \varphi_1 \vee \varphi_2 & \text{iff} & K,\Pi,i\models \varphi_1 \text{ or } K,\Pi,i\models\varphi_2, \\
  K,\Pi,i\models\LTLcircle\varphi & \text{iff} & K,\Pi,i+1\models \varphi, \\
  K,\Pi,i\models\varphi_1\U\varphi_2 & \text{iff} & \exists k \geq i :~ K,\Pi,k \models \varphi_2~\text{and} \\
  &&\forall i \leq j < k:~ K,\Pi,j\models\varphi_1, \\
  K,\Pi,i\models\exists\pi.\; \varphi & \text{iff} & \exists p \in \Paths(K,\Pi(\varepsilon)(i)) :~\\
	&& K, \Pi[\pi \mapsto p,~\varepsilon\mapsto p],i \models \varphi,
\end{array}
\]
where $\epsilon$ is a special path variable that 
denotes the path  most recently added to $\Pi$ (i.e., closest in scope to $\pi$). 
For the empty assignment $\Pi_\emptyset$, we define $\Pi_\emptyset(\varepsilon)(i)$ to yield the initial state. 
A Kripke structure $K=(S,s_0,\delta,\ap,L)$ satisfies a HyperCTL$^*$ formula $\varphi$, denoted with $K\models\varphi$, iff $K, \Pi_\emptyset\models \varphi$.

\paragraph{Expressiveness.} HyperCTL$^*$ can express the flow of information that appears in different branches of the computation tree. Consider, for example, the following Kripke structure  (taken from \cite{journals/eatcs/Finkbeiner17}):

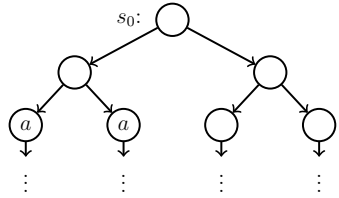
\begin{center}
\begin{tikzpicture}[level distance=7mm,line width=0.75pt,
    level 1/.style={sibling distance=26mm,
      edge from parent/.style={draw,->,solid,line width=0.75pt}},
    level 2/.style={sibling distance=13mm,
      edge from parent/.append style={draw,->,solid,line width=0.75pt}},
    level 3/.style={sibling distance=10mm,
      edge from parent/.append style={draw,->,solid,line width=0.75pt}}]
		\tikzstyle{every node}=[font=\large]
    \node [scale=0.7] (a) {$s_0$:\qquad\quad~~~~}  
				node [circle,draw,scale=0.7] (a) {\phantom{$a$}} 
        child {node [circle,draw,scale=0.7] (a1) {\phantom{$a$}}
          child {node [circle,draw,scale=0.7] (a11) {$a$}
          child {node [scale=0.7] (a1111) {\vdots}}}
          child {node [circle,draw,scale=0.7] (a21) {$a$}
            child {node [scale=0.7] (a1111) {\vdots}}}
      }
        child {node [circle,draw,scale=0.7] (a1) {\phantom{$a$}}
          child {node [circle,draw,scale=0.7] (a11) {\phantom{$a$}}
            child {node [scale=0.7] (a1111) {\vdots}}}
          child {node [circle,draw,scale=0.7] (a21) {\phantom{$a$}}
            child {node [scale=0.7] (a1111) {\vdots}}}
	};
\end{tikzpicture}
\end{center}

An observer who sees $a$ can infer which branch was taken in the first nondeterministic choice, but not which branch was taken in the second nondeterministic choice. This is expressed by the HyperCTL$^*$ formula \[\forall \pi.\, \X \forall \pi' .\, \X (a_\pi \equ a_{\pi'}).\]

\paragraph{Model checking.} The modification to the model checking algorithm from Section~\ref{section:modelchecking} needed to take care of branching time is to change to alphabet of the automata from $(2^\ap)^n$, i.e., tuples of sets of atomic propositions, to $S^n$, i.e., tuples of states of the Kripke structure. The model checking algorithm is described in detail in~\cite{DBLP:conf/cav/FinkbeinerRS15}.
The algorithm again starts by translating the inner LTL formula $\varphi$ of the negated specification into an equivalent B\"uchi automaton $\mathcal A_0$ over the alphabet $(2^\ap)^n$; this automaton is then translated into an automaton over alphabet $S^n$ by applying the labeling function $L$ to the individual positions of the tuple. 
The algorithm then proceeds as described in Section~\ref{section:modelchecking}, eliminating in each step one path quantifier. In the elimination of the quantifier, the automaton is combined as before with the Kripke structure, ensuring that the state sequence corresponds to a path in the Kripke structure.
After $n$ steps, all quantifiers have been eliminated, and the language of the resulting automaton is, as before. over the one-letter alphabet (consisting of the empty tuple). The HyperCTL$^*$ formula is satisfied if and only if the language of the resulting automaton is empty.

\section{HyperQCTL$^*$}

HyperQCTL$^*$~\cite{DBLP:conf/lics/CoenenFHH19} extends HyperCTL$^*$ with quantification over atomic propositions.
The formulas of Hyper\-QCTL$^*$ are generated by the following grammar:
\[
\begin{array}{llllllllllllllllll}
\varphi ~&~::= ~&~ a_{\pi} ~&~ \vert ~&~ \neg\varphi ~&~ \vert ~&~ 
\varphi\vee\varphi ~&~ \vert ~&~ \LTLcircle\varphi ~&~ \vert ~&~ \varphi \U\varphi ~&~ \vert ~&~ \exists \pi.\; \varphi ~&~ \vert ~&~ \boldsymbol{\exists} p.\; \varphi
\end{array}
\]
where $a,p \in \ap$ and $\pi \in \var$.
The semantics of HyperQCTL$^*$ corresponds to the semantics of HyperCTL$^*$ with an additional rule for propositional quantification. In QPTL, a propositional quantifier over a proposition $p$ determines a sequence in $(2^{p})^\omega$; i.e., the value of the proposition depends on the position in the sequence. In HyperQCTL$^*$, the quantification modifies the interpretation on the entire computation tree.
\begin{alignat*}{3}
&K, \Pi, i \models \exists q. \varphi &&\text{ iff } && \exists L': S^* \rightarrow 2^{AP \cup \set{q}}.~ \forall w \in S^*. \\
  & && && L'(w) =_{AP \setminus \set{q}} L(w) \land K[L' / L], \Pi, i \models \varphi.
\end{alignat*}
We say that a Kripke structure $K$ satisfies a HyperQCTL$^*$ formula $\varphi$, written $K \models \varphi$, if $K, \emptyset, 0 \models \varphi$.

\paragraph{Expressiveness.} HyperQCTL$^*$ is strictly more expressive than HyperCTL$^*$. In particular, HyperQCTL$^*$ subsumes the extension of HyperCTL$^*$ with the knowledge operator. The formula $K_{A,\pi} \varphi$ states that the agent who can observe the propositions $A \subseteq \ap$ on path $\pi$ knows that $\varphi$ holds. The semantics of $K_{A,\pi}$ is defined (analogously to the linear-time version in Section~\ref{section:knowledgeop}) as follows:
\begin{alignat*}{3}
  &K, \pathassign, i \models \mathcal{K}_{A,\pi} \varphi \quad &&\text{iff} && \quad \forall p \in \Paths(K,s_0) \ldot \pathassign(\pi)[0, i] =_A p[0, i] \rightarrow\\
  &&&&& \quad T, \pathassign[\pi \mapsto p], i \models_K \varphi
\end{alignat*}
HyperQCTL$^*$ also has the same expressiveness as second-order modadic logic equipped with the equal-level predicate (MSO[$E$]), i.e., the extension of \foe (as defined in Section 3) with second-order quantification~\cite{DBLP:conf/lics/CoenenFHH19}.

\paragraph{Model checking.} The model checking problem of HyperQCTL$^*$ is undecidable~\cite{DBLP:conf/lics/CoenenFHH19}.

\section{Monadic Path Logic}

Monadic path logic equipped with the equal-level predicate (\mple)
is the extension of \foe (as defined in Section~\ref{section:foe}) with second-order quantification, where the second-order quantification is restricted to full paths in the Kripke structure.

Let $V_1 = \{x_1, x_2, \ldots\}$ be a set of first-order variables, and $V_2 = \{X_1, X_2, \ldots\}$ a set of second-order variables.
The formulas $\varphi$ of \mple are generated by the following grammar: 
\begin{align*}	
  \varphi &\Coloneqq \psi ~|~ \neg \varphi ~|~ \varphi_1 \vee \varphi_2  ~|~ \exists x. \varphi ~|~ \exists X. \varphi\\
\psi  &\Coloneqq P_a(x) ~|~ x<y ~|~ x=y ~|~ x\in X ~|~ E(x,y), 
\end{align*}
where $a \in AP$, $x, y \in V_1$, and $X \in V_2$.
In the semantics of \mple, we assign first-order variables to sequences of states that form a prefix of a path in the Kripke structure, and second-order variables to the infinite prefix-closed sets of prefixes of the paths of the Kripke structure.

We define the satisfaction relation $K, \Veins, \Vzwei \models \varphi $ for a Kripke structure $K$ and two valuations $\mathcal V_1, \mathcal V_2$ as follows:
\begin{align*}
  K, \Veins, \Vzwei &\models P_a(x)&&\text{ iff } && a \in L(\Veins(x))\\
  K, \Veins, \Vzwei &\models x<y&&\text{ iff } && \Veins(x) \sqsubseteq \Veins(y)   \\
  K, \Veins, \Vzwei &\models x=y&&\text{ iff } && \Veins(x) = \Veins(y)   \\
  K, \Veins, \Vzwei &\models x \in X&&\text{ iff } && \Veins(x) \in \Vzwei(X)   \\
  K, \Veins, \Vzwei &\models E(x,y)&&\text{ iff } && |\Veins(x)| = |\Veins(y)|\\
  K, \Veins, \Vzwei &\models \neg \varphi &&\text{ iff } &&  K, \Veins, \Vzwei \not\models \varphi\\
  K, \Veins, \Vzwei &\models \varphi_1 \vee \varphi_2 &&\text{ iff } && K, \Veins, \Vzwei \models \varphi_1 \lor  K, \Veins, \Vzwei \models \varphi_2 \\
  K, \Veins, \Vzwei &\models  \exists x. \varphi &&\text{ iff } && \exists p \in S^*, p' \in \Paths(K, s_0).\ p \sqsubseteq p' \land\\
  &&&&& K, \Veins[x \mapsto p], \Vzwei \models \varphi\\
  K, \Veins, \Vzwei &\models \exists X. \varphi &&\text{ iff } && \exists p \in \Paths(K, s_0).\ \Veins, \Vzwei[X \mapsto \mathit{Prefixes}(p)] \models \varphi
\end{align*}
where $\mathcal{V}_i[x \mapsto v]$ updates a valuation, $p_1 \sqsubseteq p_2$ denotes that $p_1$ is a prefix of $p_2$, and $\mathit{Prefixes}(p)$ is the set of prefixes of $p$.
A Kripke structure $K$ satisfies a closed \mple formula $\varphi$, written $K \models \varphi$, if $T, \emptyset, \mathcal{V}_2 \models \varphi$, where $\emptyset$ denotes the empty first-order valuation and $\mathcal{V}_2$ assigns each free $X_a$ in $\varphi$ to the set $\{p \in S^* ~|~ a \in L(p)\}$.

\paragraph{Expressiveness.} The expressiveness of \mple falls strictly between HyperCTL$^*$ and HyperQCTL$^*$. Like HyperQCTL$^*$, \mple can, however, express the properties of HyperKCTL$^*$, i.e., the extension of HyperCTL$^*$ with the knowledge operator~\cite{DBLP:conf/lics/CoenenFHH19}.

\paragraph{Model checking.} Similar to the model checking algorithm of Section~\ref{section:modelchecking}, we reduce the model checking problem of \mple to the language emptiness problem of a B\"uchi automaton. Let $\varphi$ be the negation of the given formula. We translate $\varphi$ into an automaton $\mathcal A$ over the tuple alphabet $(S\cup\{\bot\})^{\Veins \cup \Vzwei}$ such that the language of $\mathcal A$ is empty iff the original formula is satisfied by the Kripke structure.
The automaton is constructed recursively as follows:
\begin{itemize}
\item If $\varphi= P_a(x)$, then $\mathcal A$ accepts all infinite sequences where the first time the component of component of $x$ becomes $\bot$ at some point, and stays $\bot$ from thereon after, and $a$ is contained in $L(w)$ where $w$ is the sequence of states in $x$'s component up to that point.
\item If $\varphi = x<y$, then $\mathcal A$ accepts all infinite sequences where the components of $x$ and $y$ each become and stay $\bot$ at some point, and until $x$ becomes $\bot$ the components are the same.
\item If $\varphi = x{=}y$, then $\mathcal A$ accepts all infinite sequences where the components of $x$ and $y$ each become and stay $\bot$ at the same point, and until then the components are the same.
\item If $\varphi = x \in X$, then $\mathcal A$ accepts all infinite sequences where the component of $x$ becomes and stays $\bot$ at some point, and until then the components of $x$ and $X$ are the same.
\item If $\varphi= E(x,y)$, then $\mathcal A$ accepts all infinite sequences where the components of $x$ and $y$ become and stay $\bot$ at the same point.
\item If $\varphi = \neg \psi$, then we first compute and negate the automaton for $\psi$. $\mathcal A$ is then the intersection of that automaton with an automaton that ensures that, for every $x \in \Veins$, the component of $x$ eventually becomes and stays $\bot$.
\item If $\varphi = \exists x. \psi$, then we first compute the automaton for $\psi$. We then combine the automaton with the Kripke structure to ensure that the component for $x$ forms a prefix of a path in the Kripke structure and ends in $\bot$. $\mathcal A$ is then the existential projection of that automaton, where the component for $x$ is eliminated.
\item If $\varphi = \exists X. \psi$, then we also first compute the automaton for $\psi$. We then combine the automaton with the Kripke structure to ensure that the component for $X$ forms a full path in the Kripke structure. $\mathcal A$ is then the existential projection of that automaton, where the component for $X$ is eliminated.
  
\end{itemize}

\section{Conclusions}

We have studied the hierarchy of hyperlogics from the perspective of the model checking problem. For the logics considered here, HyperQPTL is clearly the most interesting linear-time logic, because it can still be checked using the basic model checking algorithm, while for more expressive logics like HyperQPTL$^+$ and \seins the model checking problem is already undecidable.
Among the branching-time logics, \mple has a similar position, more expressive than HyperCTL$^*$, but, unlike HyperQCTL$^*$, still with a decidable model checking problem.

From a practical point of view, the key challenge that needs to be addressed in all these logics is the treatment of quantifier alternations. In the model checking algorithm quantifier alternations lead to alternations between existential and universal projection on the constructed automaton. Such alternations can in theory be implemented using complementation; in practice, however, the exponential cost of complementation is too expensive. Model checking implementations like MCHyper therefore instead rely on quantifier elimination via strategies~\cite{DBLP:conf/cav/CoenenFST19}. In this approach, the satisfaction of a  formula of the form $\varphi = \forall \pi_1. \exists \pi_2. \psi$ is analyzed as a game between a \emph{universal} player, who chooses $\pi_1$, and an \emph{existential} player, who chooses $\pi_2$. The formula $\varphi$  is satisfied if the existential player has a strategy that ensures that $\psi$ becomes true.

\section*{Acknowledgements}
Most of the work reported in this paper has previously appeared in various publications~\cite{DBLP:conf/post/ClarksonFKMRS14,DBLP:conf/lics/CoenenFHH19,DBLP:conf/cav/CoenenFST19,10.1007/978-3-030-53291-8_4,DBLP:conf/cav/FinkbeinerRS15,Martin}. 
I am indebted to my coauthors Michael R. Clarkson, Norine Coenen, Christopher Hahn, Jana Hofmann, Masoud Koleini, Kristopher K. Micinski, Markus N. Rabe, C\'esar S\'anchez, Leander Tentrup, and Martin Zimmermann.
This work was partially supported by the Collaborative Research Center “Foundations of Perspicuous Software Systems” (TRR: 248, 389792660) and the European Research Council (ERC) Grant OSARES (No. 683300).

\bibliographystyle{plain}
\bibliography{biblio,bibliography}
\end{document}